\documentclass[aps]{revtex4}
\usepackage{amssymb,epsf}

\begin{document}

\title{Thermodynamic instability of topological black holes with nonlinear
source}
\author{S. H. Hendi$^{1,2}$\footnote{email address: hendi@shirazu.ac.ir} and M. Momennia$^{1}$}
\affiliation{$^1$ Physics Department and Biruni Observatory, College of Sciences, Shiraz
University, Shiraz 71454, Iran\\
$^2$ Research Institute for Astronomy and Astrophysics of Maragha (RIAAM),
Maragha, Iran}

\begin{abstract}
In this paper, we obtain higher dimensional topological black hole solutions
of Einstein-$\Lambda$ gravity in the presence of a class of nonlinear
electrodynamics. First, we calculate the conserved and thermodynamic
quantities of ($n+1$)-dimensional asymptotically flat solutions and show
that they satisfy the first law of thermodynamics. Also, we investigate the
stability of these solutions in the (grand) canonical ensemble. Second, we
endow a global rotation to the static Ricci-flat solutions and calculate the
conserved quantities of solutions by using the counterterm method. We obtain
a Smarr-type formula for the mass as a function of the entropy, the angular
momenta and the electric charge, and show that these quantities satisfy the
first law of thermodynamics. Then, we perform a stability analysis of the
rotating solutions both in the canonical and the grand canonical ensembles.
\end{abstract}

\maketitle

%%%%%%%%%%%%%%%%%%%%%%%%%%%%%%%%%%%%%%%%%%%%%%%%%%%%%%%%%%%%%%%%%%%%%%%%%%%%%%%%%%%%%%%%%%%%%%%%%%%%%%%%%%%%%%%%%%%%%%%%%%%%%%%%%

\section{Introduction}

Nonlinear field theories are of interest to different branches of
mathematical physics because most physical systems are inherently
nonlinear in the nature. The main reason to consider the nonlinear
electrodynamics (NLED) comes from the fact that these theories are
considerably richer than the Maxwell field and in special case
they reduce to the linear Maxwell theory. Various limitations of
the Maxwell theory, such as description of the self-interaction of
virtual electron-positron pairs \cite{H-E,Schwinger} and the
radiation propagation inside specific materials \cite{NLmaterial},
motivate one to consider NLED \cite{Delphenich}. Besides, NLED
improves the basic concept of gravitational redshift and its
dependency of any background magnetic field as compared to the
well-established method introduced by standard general relativity.
In addition, it was recently shown that NLED
objects can remove both of the big bang and black hole singularities \cite%
{bigbang}. Moreover, from astrophysical point of view, one finds that the
effects of NLED become indeed quite important in superstrongly magnetized
compact objects, such as pulsars and particular neutron stars (also the
so-called magnetars and strange quark magnetars) \cite{magnetars}.

About eighty years ago Born and Infeld introduced an interesting
kind of NLED in order to remove the divergence of the self-energy
of a point-like charge \cite{Born}. The first attempt to couple
the NLED with gravity was made by Hoffmann \cite{Hoffmann}. After
that the effects of Born-Infeld (BI) NLED coupled to the
gravitational field have been studied for static black holes
\cite{Blackhole}, rotating black objects \cite{Rotating},
wormholes \cite{Wormhole,Worm} and superconductors
\cite{Superconductor}. Also, BI NLED has acquired a new impetus,
since it naturally arises in the low-energy limit of the open
string theory \cite{Stringtheory}. Recently, two different BI
types of NLED have been introduced, which can also remove the
divergence of the electric field near the origin. One of them is
Soleng NLED which is logarithmic form \cite{Soleng} and another
one was proposed by Hendi with exponential form \cite{Hendi}. The
Soleng field, like BI theory, removes divergency of the electric
field, while the theory proposed by Hendi does not. It is notable
to mention that although the exponential form of NLED does not
cancel the divergency of the electric field but its singularity is
much weaker than that in the Maxwell theory. Black object
solutions coupled to these two
nonlinear fields have been studied in literature (for e.g., see \cite%
{NEs,Worm}). The Lagrangian of mentioned BI type nonlinear theories, for
weak nonlinearity, tends to the following form
\begin{equation}
L(\mathcal{F})=-\mathcal{F}+\alpha \mathcal{F}^{2}+O\left( \alpha
^{2}\right) ,  \label{Lagrangian}
\end{equation}%
where $\mathcal{F}=F_{\mu \nu }F^{\mu \nu }$ is the Maxwell invariant, in
which $F_{\mu \nu}=\partial _{\mu}A_{\nu}-\partial _{\nu}A_{\mu}$ is the
electromagnetic field tensor and $A_{\mu}$ is the gauge potential. In
addition, $\alpha$ denotes nonlinearity parameter which is small and so the
effects of nonlinearity should be considered as a perturbation ($\alpha$ is
proportional to the inverse value of nonlinearity parameter in BI-type
theories). In this paper, we take into account the Eq. (\ref{Lagrangian}) as
a NLED source and investigate the effects of nonlinearity on the properties
of static and rotating black hole/brane solutions.

Here, it is necessary to focus on the basic motivation of considering the
Lagrangian (\ref{Lagrangian}). At first, we should note that, regardless of
a constant parameter, most of NLED Lagrangians reduce to Eq. (\ref%
{Lagrangian}) for the weak nonlinearity. Eventually, it is worthwhile to
mention that although various theories of NLED have been created with
different primitive motivations, only for the weak nonlinearity (Eq. (\ref%
{Lagrangian})), they contain physical and experimental importances. As we
know, using the Maxwell theory in various branches leads to near accurate or
acceptable consequences. So, in transition from the Maxwell theory to NLED,
the logical decision is to consider the effects of weak nonlinearity
variations, not strong effects. This means that, one can expect to obtain
precise physical results with experimental agreements, provided one regards
the nonlinearity as a correction to the Maxwell field.

For the reasons mentioned above, there have been published some reasonable
works by considering Eq. (\ref{Lagrangian}) as an effective Lagrangian of
electrodynamics \cite%
{H-E,Schwinger,Stehle,Delphenich,Kats,BIString,Frad,Fradkin}. Heisenberg and
Euler have shown that quantum corrections lead to nonlinear properties of
vacuum \cite{H-E,Schwinger,Stehle,Delphenich}. Also, as we mentioned before,
it was proved that in the low-energy limit of heterotic string theory, a
quartic correction of the Maxwell field strength tensor appears \cite%
{Kats,BIString,Frad,Fradkin}. So it is natural to consider Eq. (\ref%
{Lagrangian}) as an effective and suitable Lagrangian of
electrodynamics instead of the Maxwell one.

The outline of our paper is as follows. In the next section, we consider the
$(n+1)$-dimensional topological static black hole solutions of Einstein
gravity in presence of the mentioned NLED and investigate their properties.
In Sec. \ref{K1}, we calculate the conserved and thermodynamic quantities of
asymptotically flat black holes, check the first law of thermodynamics and
investigate the stability of the solutions in both canonical and grand
canonical ensembles. Sec. \ref{K0} is devoted to introducing the rotating
solutions with flat horizon and computing the conserved and thermodynamic
quantities of the solutions. We also check the first law of thermodynamics
and perform the stability analysis of the solutions both in the canonical
and the grand canonical ensembles for the rotating solutions. We finish our
paper with some concluding remarks.

%%%%%%%%%%%%%%%%%%%%%%%%%%%%%%%%%%%%%%%%%%%%%%%%%%%%%%%%%%%%%%%%%%%%%%%%%%%%%%%%%%%%%%%%%%%%%%%%%%%%%%%%%%%%%%%%%%%%%%%%%%%%%%%%%

\section{Static Topological Black Hole Solutions \label{FE}}

The $(n+1)$-dimensional action of Einstein gravity with negative
cosmological constant and in presence of nonlinear electrodynamics is
\begin{equation}
I_{G}=-\frac{1}{16\pi }\int_{\mathcal{M}}d^{n+1}x\sqrt{-g}\left[ R-2\Lambda
+L(\mathcal{F})\right] -\frac{1}{8\pi }\int_{\partial \mathcal{M}}d^{n}x%
\sqrt{-\gamma }\Theta \left( \gamma \right) ,  \label{Action}
\end{equation}
where $R$ is the scalar curvature, $\Lambda$ is the cosmological constant
which is equal to $-n\left( n-1\right) /2l^{2}$ for asymptotically adS
solutions. In this action, $L(\mathcal{F})$ is the Lagrangian of nonlinear
electrodynamics presented in Eq. (\ref{Lagrangian}) and the second integral
is the Gibbons-Hawking surface term which is chosen such that the
variational principle will be well defined \cite{Myers}. In the second
integral, $\gamma$ and $\Theta $ are, respectively, the trace of induced
metric, $\gamma _{ij}$, and the extrinsic curvature $\Theta _{ij}$ on the
boundary $\partial \mathcal{M}$. Variation of the action (\ref{Action}) with
respect to the metric tensor $g_{\mu \nu }$ and the Faraday tensor $F_{\mu
\nu}$, leads to
\begin{equation}
G_{\mu \nu }+\Lambda g_{\mu \nu }=\frac{1}{2}g_{\mu \nu }L(\mathcal{F})-2L_{%
\mathcal{F}}F_{\mu \lambda }F_{\nu }^{\lambda },  \label{Field equation}
\end{equation}

\begin{equation}
\partial _{\mu }\left( \sqrt{-g}L_{\mathcal{F}}F^{\mu \nu }\right) =0,
\label{Maxwell equation}
\end{equation}
where $G_{\mu \nu }$ is the Einstein tensor and $L_{\mathcal{F}}=dL(\mathcal{%
F)}/d\mathcal{F}$.

Here, we want to obtain the $(n+1)$-dimensional topological static black
hole solutions. We take into account the metric of $(n+1)$-dimensional
spacetime with the following form
\begin{equation}
ds^{2}=-f(r)dt^{2}+\frac{dr^{2}}{g(r)}+r^{2}d\Omega _{n-1}^{2},
\label{metric}
\end{equation}%
where $f(r)$\ and $g(r)$ are two arbitrary functions of radial coordinate
which should be determined and
\begin{equation}
d\Omega _{n-1}^{2}=\left\{
\begin{array}{cc}
d\theta _{1}^{2}+\sum\limits_{i=2}^{n-1}\prod\limits_{j=1}^{i-1}\sin
^{2}\theta _{j}d\theta _{i}^{2} & k=1 \\
d\theta _{1}^{2}+ \sinh ^{2}\theta _{1} \left(d\theta
_{2}^{2}+\sum\limits_{i=3}^{n-1}\prod\limits_{j=2}^{i-1}\sin ^{2}\theta
_{j}d\theta _{i}^{2}\right) & k=-1 \\
\sum\limits_{i=1}^{n-1}d\phi _{i}^{2} & k=0%
\end{array}%
\right. ,  \label{dOmega}
\end{equation}%
represents the line element of an $\left( n-1\right) $-dimensional
hypersurface with constant curvature $\left( n-1\right) \left( n-2\right) k$
and volume $V_{n-1}$. We should note that the constant $k$ characterizes the
hypersurface and indicates that the boundary of $t=constant$ and $r=constant$
can be a positive (elliptic), zero (flat) or negative (hyperbolic) constant
curvature hypersurface.

%%%%%%%%%%%%%%%%%%%%%%%%%%%%%%%%  Metric1  %%%%%%%%%%%%%%%%%%%%%%%%%%
\begin{figure}[tbp]
\epsfxsize=10cm \epsffile{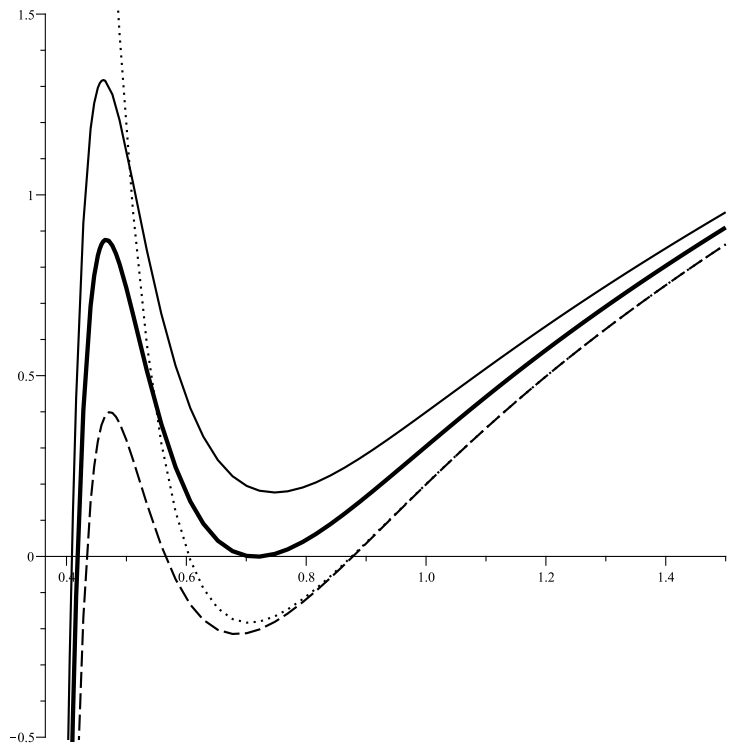}
\caption{\textbf{elliptic horizon solutions:} $f(r)$ versus $r$ for $k=1$, $%
n=4$, $q=1$, $\Lambda=-1$, $\protect\alpha=0.005$, and $m=1.1$ (solid line),
$m=1.2$ (bold line) and $m=1.3$ (dashed line); "dotted line is $f(r)$ for
the Maxwell case ($\protect\alpha=0$) with $m=1.3$ ".}
\label{metric1}
\end{figure}
%%%%%%%%%%%%%%%%%%%%%%%%%%%%%%%%%%%%%%%%%%%%%%%%%%%%%%%%%%
%%%%%%%%%%%%%%%%%%%%%%%%%%%%%%%%  Metric2  %%%%%%%%%%%%%%%%%%%%%%%%%%
\begin{figure}[tbp]
\epsfxsize=10cm \epsffile{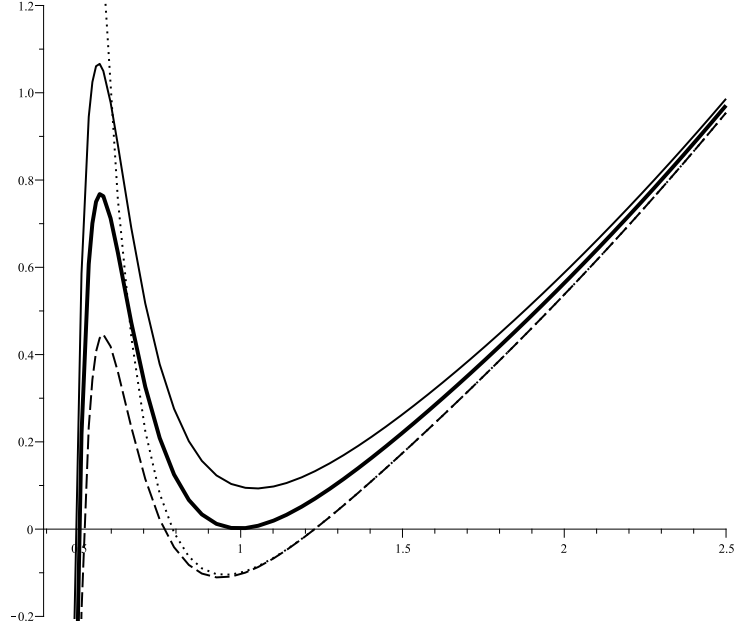}
\caption{\textbf{flat horizon solutions:} $f(r)$ versus $r$ for $k=0$, $n=4$%
, $q=1$, $\Lambda=-1$, $\protect\alpha=0.02$, and $m=0.4$ (solid line), $%
m=0.5$ (bold line) and $m=0.6$ (dashed line) ; "dotted line is $f(r)$ for
the Maxwell case ($\protect\alpha=0$) with $m=0.6$ ".}
\label{metric2}
\end{figure}
%%%%%%%%%%%%%%%%%%%%%%%%%%%%%%%%%%%%%%%%%%%%%%%%%%%%%%%%%%
%%%%%%%%%%%%%%%%%%%%%%%%%%%%%%%%  Metric3  %%%%%%%%%%%%%%%%%%%%%%%%%%
\begin{figure}[tbp]
\epsfxsize=10cm \epsffile{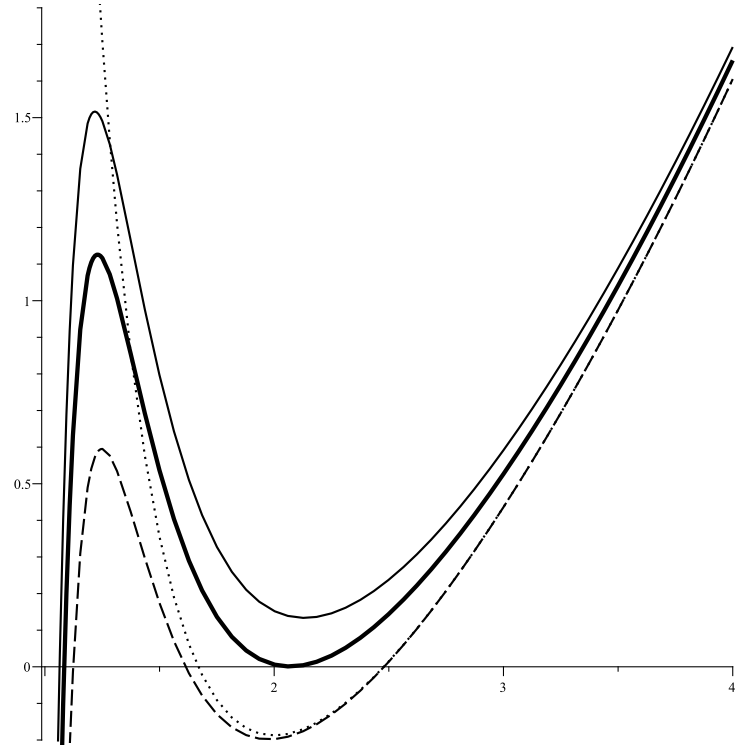}
\caption{\textbf{hyperbolic horizon solutions:} $f(r)$ versus $r$ for $k=-1$%
, $n=4$, $q=5$, $\Lambda=-1$, $\protect\alpha=0.1$, and $m=0.1$ (solid
line), $m=0.7$ (bold line) and $m=1.5$ (dashed line); "dotted line is $f(r)$
for the Maxwell case ($\protect\alpha=0$) with $m=1.5$ " .}
\label{metric3}
\end{figure}
%%%%%%%%%%%%%%%%%%%%%%%%%%%%%%%%%%%%%%%%%%%%%%%%%%%%%%%%%%

%%%%%%%%%%%%%%%%%%%%%%%%%%%%%%%%  Metric4  %%%%%%%%%%%%%%%%%%%%%%%%%%
\begin{figure}[tbp]
\epsfxsize=10cm \epsffile{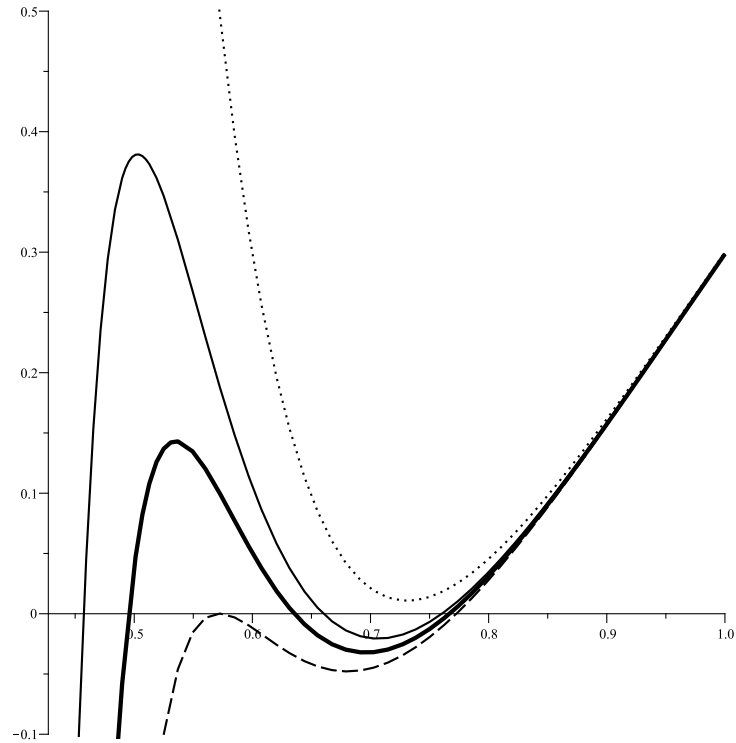}
\caption{\textbf{elliptic horizon solutions:} $f(r)$ versus $r$ for $k=1$, $%
n=4$, $q=1$, $\Lambda=-1$, $m=1.2$, and $\protect\alpha=0.007$ (solid line),
$\protect\alpha=0.009$ (bold line) and $\protect\alpha=0.011$ (dashed line);
"dotted line is $f(r)$ for the Maxwell case ($\protect\alpha=0$) with $m=1.2$%
".}
\label{metric4}
\end{figure}
%%%%%%%%%%%%%%%%%%%%%%%%%%%%%%%%%%%%%%%%%%%%%%%%%%%%%%%%%%

%%%%%%%%%%%%%%%%%%%%%%%%%%%%%%%%  Penrose1  %%%%%%%%%%%%%%%%%%%%%%%%%%
\begin{figure}[tbp]
$%
\begin{array}{cc}
\epsfxsize=8cm \epsffile{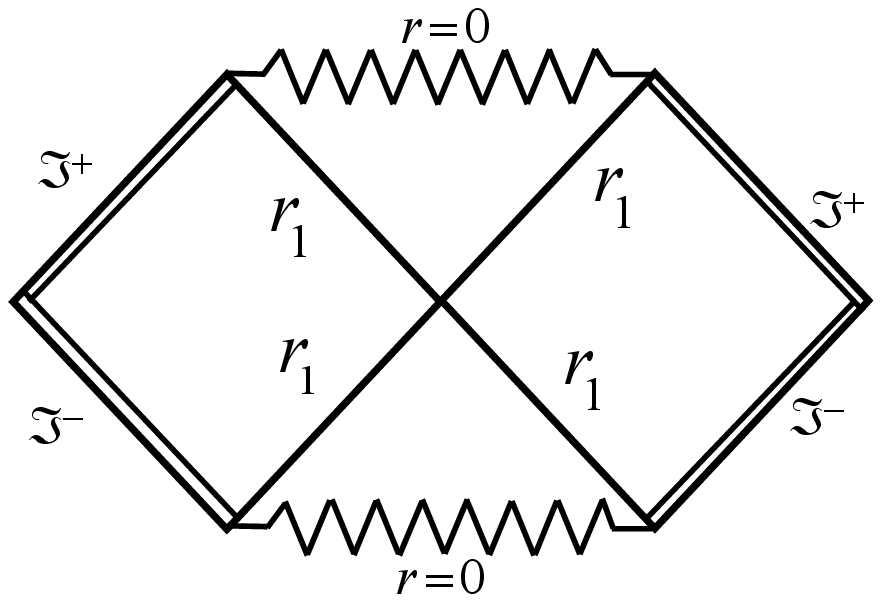} & \epsfxsize=6cm %
\epsffile{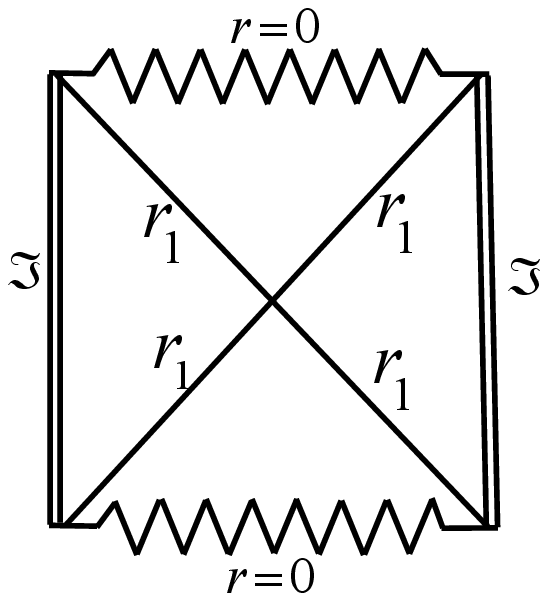}%
\end{array}
$%
\caption{ Carter-Penrose diagram for the asymptotically flat (left figure)
and the asymptotically adS (right figure) black holes when the metric
function has one real positive root ($r_{1}$) (the same as Schwarzschild
black hole). }
\label{Pen1}
\end{figure}
%%%%%%%%%%%%%%%%%%%%%%%%%%%%%%%%%%%%%%%%%%%%%%%%%%%%%%%%%%

%%%%%%%%%%%%%%%%%%%%%%%%%%%%%%%%  Penrose2  %%%%%%%%%%%%%%%%%%%%%%%%%%
\begin{figure}[tbp]
$%
\begin{array}{cc}
\epsfxsize=8cm \epsffile{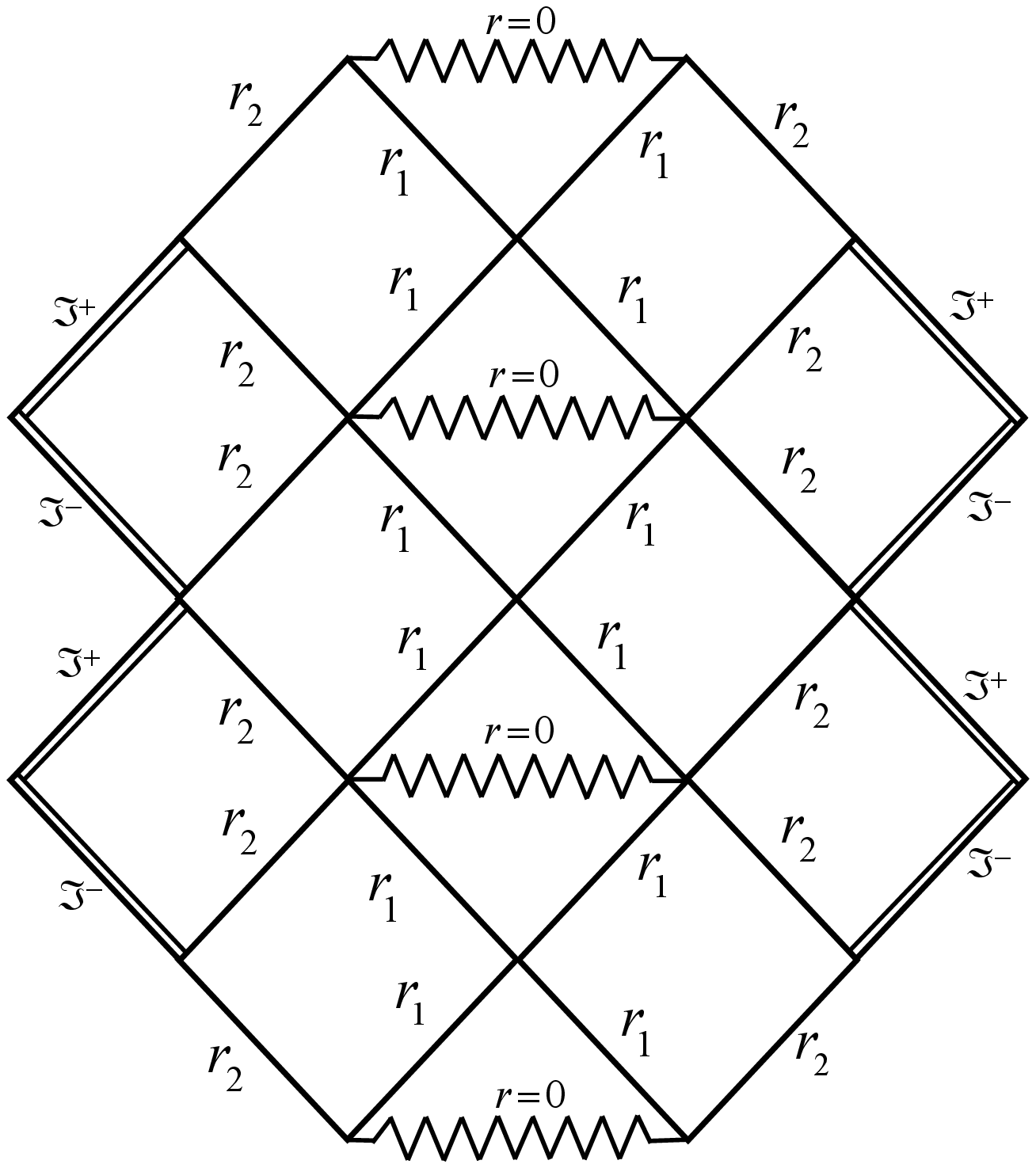} & \epsfxsize=6cm %
\epsffile{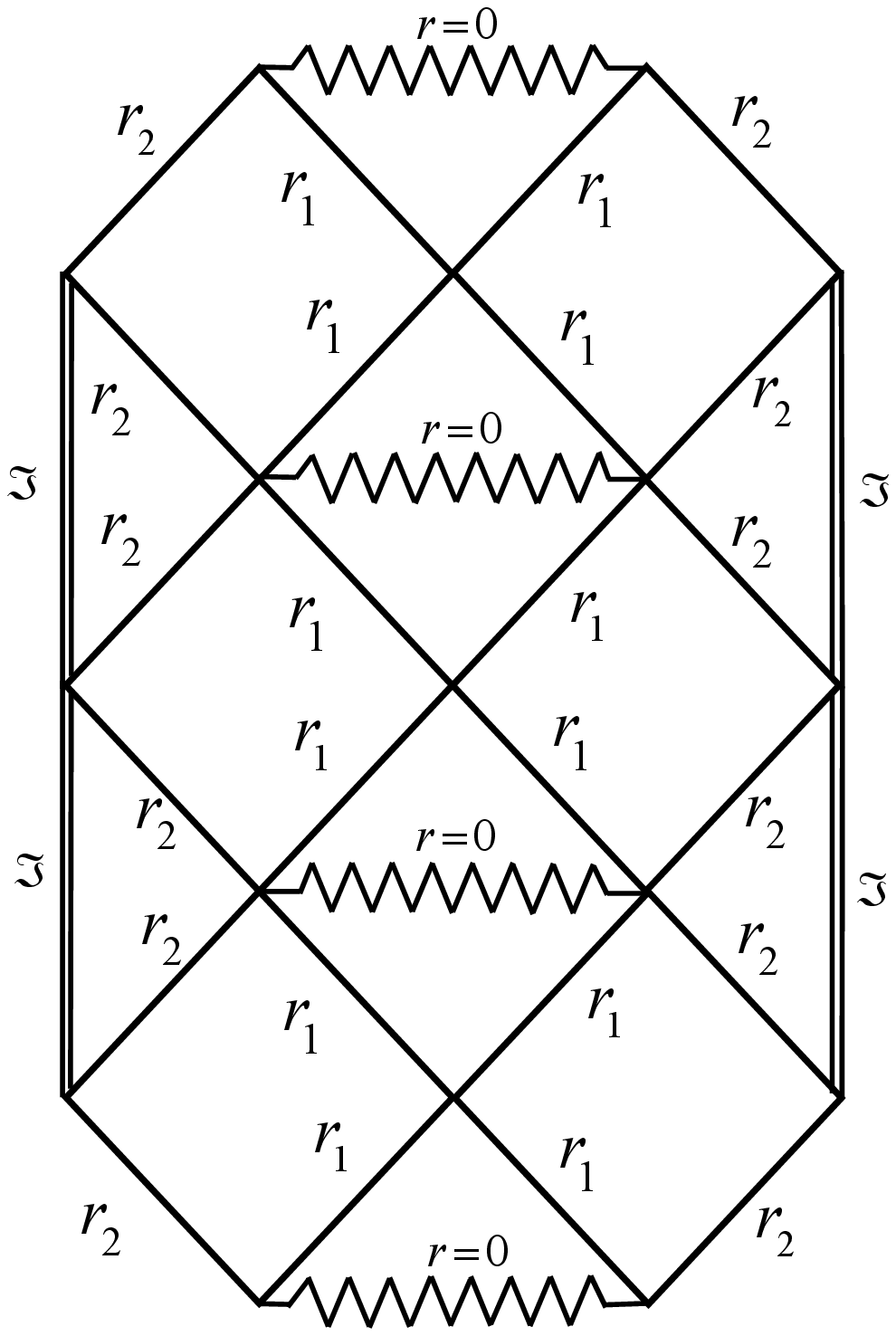}%
\end{array}
$%
\caption{Carter-Penrose diagram for the asymptotically flat (left figure)
and the asymptotically adS (right figure) black holes when the metric
function has two real positive roots ($r_{1}$ and $r_{2}$) (second root is
an extreme root). }
\label{Pen2}
\end{figure}
%%%%%%%%%%%%%%%%%%%%%%%%%%%%%%%%%%%%%%%%%%%%%%%%%%%%%%%%%%

%%%%%%%%%%%%%%%%%%%%%%%%%%%%%%%%  Penros3  %%%%%%%%%%%%%%%%%%%%%%%%%%
\begin{figure}[tbp]
$%
\begin{array}{cc}
\epsfxsize=6cm \epsffile{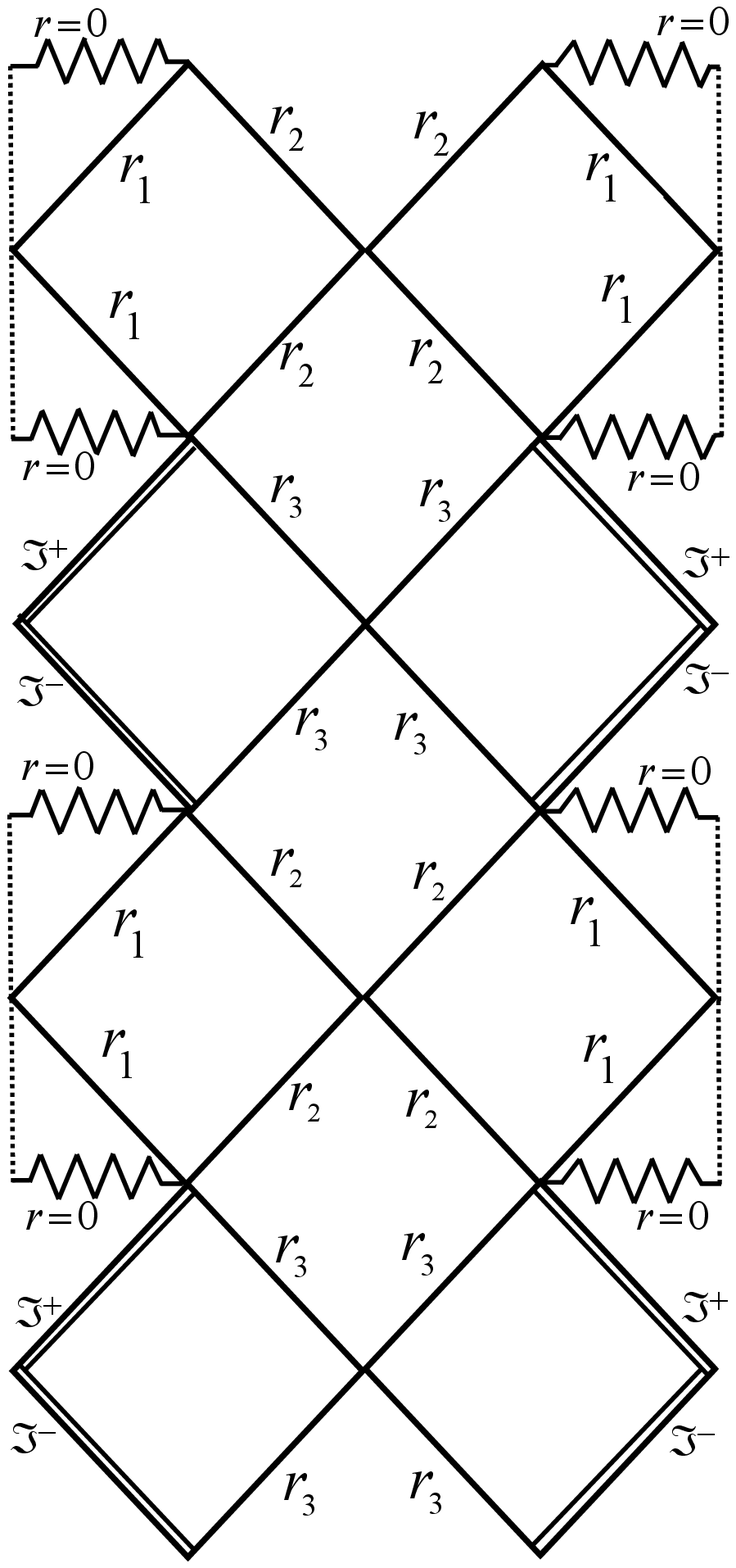} & \epsfxsize=6cm %
\epsffile{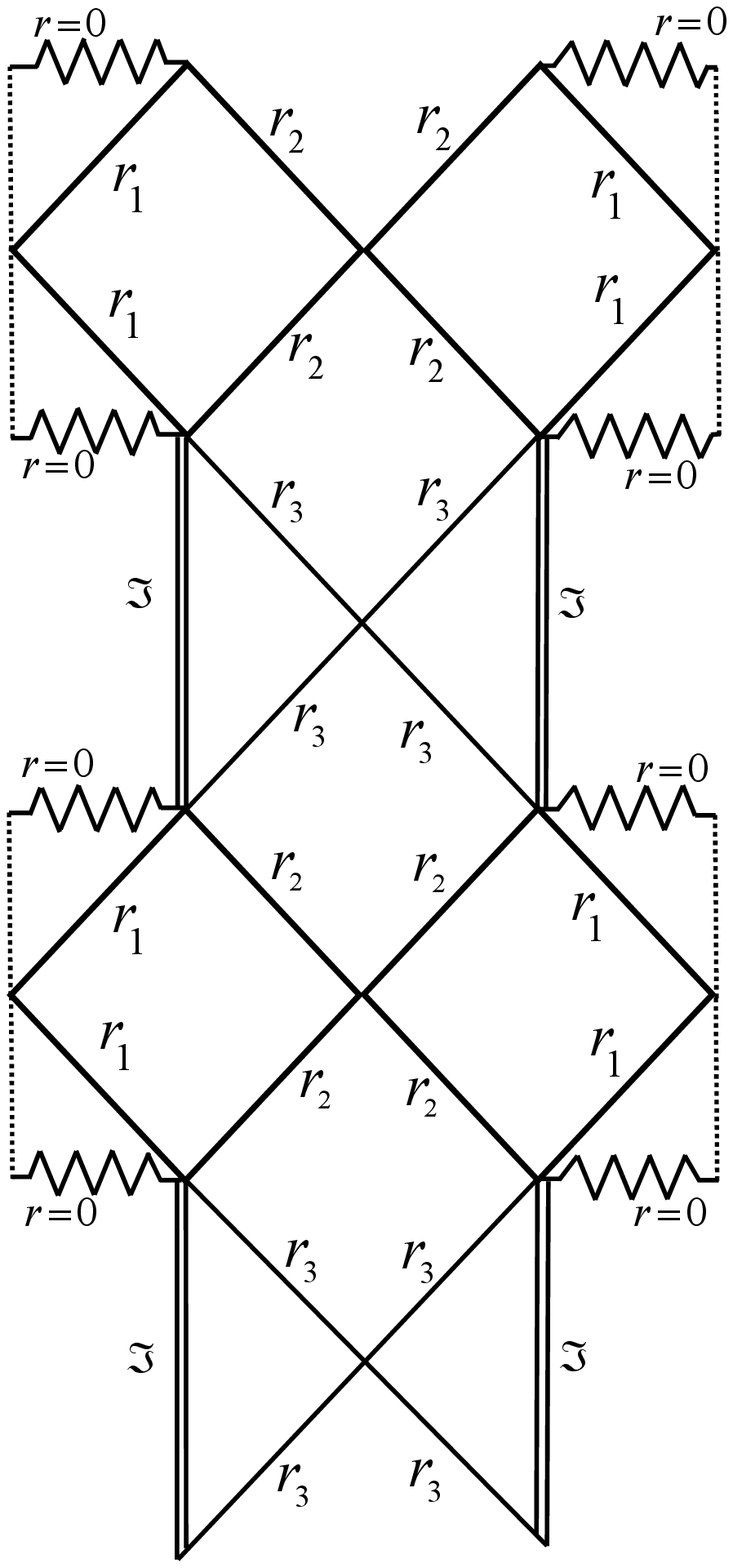}%
\end{array}
$%
\caption{Carter-Penrose diagram for the asymptotically flat (left figure)
and the asymptotically adS (right figure) black holes when the metric
function has three real positive roots ($r_{1}$, $r_{2}$ and $r_{3}$). }
\label{Pen3}
\end{figure}
%%%%%%%%%%%%%%%%%%%%%%%%%%%%%%%%%%%%%%%%%%%%%%%%%%%%%%%%%%

Using Eq. (\ref{Maxwell equation})\ with the following radial gauge
potential ansatz
\begin{equation}
A_{\mu }=h\left( r\right) \delta _{\mu }^{0},  \label{potential ansatz}
\end{equation}%
we obtain the following differential equation
\begin{equation}
E^{\prime }(r)+\frac{(n-1)E(r)}{r}+4E^{2}(r)\left( \frac{(n-1)E(r)}{r}%
+3E^{\prime }(r)\right) \alpha =0,  \label{Eeq}
\end{equation}%
where $E(r)=F_{tr}=h^{\prime }(r)$ is the nonzero component of
electromagnetic field and prime denotes the derivative with respect to $r$.
Solving Eq. (\ref{Eeq}), one obtains%
\begin{equation}
E(r)=\frac{3\left[ q\alpha ^{2}r^{2n+1}\left( 1+\sqrt{1+\frac{r^{2n-2}}{%
27\alpha q^{2}}}\right) \right] ^{2/3}-\alpha r^{2n}}{6\alpha r^{n}\left[
q\alpha ^{2}r^{2n+1}\left( 1+\sqrt{1+\frac{r^{2n-2}}{27\alpha q^{2}}}\right) %
\right] ^{1/3}},  \label{E1(r)}
\end{equation}%
where $q$ is an integration constant which is related to the electric charge
of the black hole. Now, we use the series expansion of $E(r)$ for small
values of $\alpha $, and keep the first two terms to obtain%
\begin{equation}
E(r)=\frac{q}{r^{n-1}}-4\left( \frac{q}{r^{n-1}} \right)^3 \alpha +O\left(
\left( \frac{q}{r^{n-1}} \right)^5 \alpha ^{2}\right) ,  \label{E2(r)}
\end{equation}%
or correspondingly%
\begin{equation}
h(r)=-\frac{q}{\left( n-2\right) r^{n-2}}+\frac{4q^{3}}{\left( 3n-4\right)
r^{3n-4}}\alpha +O\left( \alpha ^{2}\right) .  \label{h(r)}
\end{equation}%
It is easy to see that the second term in Eqs. (\ref{E2(r)}) and (\ref{h(r)}%
) comes from the nonlinear correction and for vanishing $\alpha $,
one can reproduce the results of the Maxwell theory. Since we want
to investigate the nonlinearity parameter as a (perturbative)
correction, hereafter, we take into account the first correction
term of nonlinearity parameter, $\alpha $, and ignore $\alpha
^{2}$ and higher power of nonlinearity parameter terms. To obtain
the metric functions $f\left( r\right) $ and $g(r)$, one may use
the nonzero components of Eq. (\ref{Field equation}).
Straightforward calculations show that the nonzero components of
Eq. (\ref{Field equation})
(up to the first order of $\alpha $) can be written as%
\begin{eqnarray}
e_{1} &=&(n-1)rf^{\prime }(r)+(n-1)(n-2)\left[ f(r)-k\right] +2\Lambda r^{2}+%
\frac{2q^{2}}{r^{2n-4}}-\frac{4q^{4}\alpha }{r^{4n-6}}+O\left( \alpha
^{2}\right) =0,  \label{e1} \\
e_{2} &=&r^{2}f^{\prime \prime }(r)+2(n-2)rf^{\prime }(r)+(n-2)(n-3)\left[
f(r)-k\right] +2\Lambda r^{2}-\frac{2q^{2}}{r^{2n-4}}+\frac{12q^{4}\alpha }{%
r^{4n-6}}+O\left( \alpha ^{2}\right) =0.  \label{e2}
\end{eqnarray}
After some calculations, one can show that the solutions of Eqs. (\ref{e1})
and (\ref{e2}) can be written as%
\begin{eqnarray}
f(r) &=&k-\frac{m}{r^{n-2}}-\frac{2\Lambda r^{2}}{n\left( n-1\right) }+\frac{%
2q^{2}}{\left( n-1\right) \left( n-2\right) r^{2n-4}}-  \nonumber \\
&&\frac{4q^{4}}{\left[ 2\left( n-2\right) \left( n+2\right) +\left(
n-3\right) \left( n-4\right) \right] r^{4n-6}}\alpha +O\left( \alpha
^{2}\right) ,  \label{metric function}
\end{eqnarray}%
\begin{equation}
g(r)=Cf(r),  \label{gf}
\end{equation}%
where $m$ is an integration constant which is related to the mass of the
black hole and the last term in Eq. (\ref{metric function}) indicates the
effect of nonlinearity. Hereafter, we set the constant $C=1$ without loss of
generality. It is notable to mention that, for $\alpha =0$, this metric
function reduces to Reissner-Nordstr\"{o}m solution, as it should. The
asymptotical behavior of the solution (\ref{metric function}) is adS or dS
provided $\Lambda <0$ or $\Lambda >0$ and the case of asymptotically flat
solutions is permitted for $\Lambda =0$ and $k=1$.

Now we look for the singularities of the solutions. One can show that the
metric (\ref{metric}) with the metric function (\ref{metric function})\ has
an essential singularity at $r=0$\ by calculating the Kretschmann scalar, as
\begin{eqnarray}
R_{\mu \nu \lambda \kappa }R^{\mu \nu \lambda \kappa } &=&\frac{8\Lambda ^{2}%
}{n(n+1)}-\frac{16\Lambda q^{2}}{n(n+1)r^{2n}}+\frac{\left( n^{2}-1\right)
\left( n^{2}-2\right) m^{2}}{r^{2n+2}}-  \nonumber \\
&&\frac{8\left( 2n^{2}-n-2\right) mq^{2}}{r^{3n+1}}+\left( \frac{96\Lambda
\alpha }{n+1}+\frac{8\left( 8n^{2}-12n+3\right) }{n-1}\right) \frac{q^{4}}{%
nr^{4n}}+  \nonumber \\
&&\frac{32\left( n-1\right) \left( 4n^{2}-n-2\right) m\alpha q^{4}}{\left(
3n-1\right) r^{5n+1}}-\frac{32\left( 32n^{2}-32n+5\right) \alpha q^{6}}{%
n\left( 3n-1\right) r^{6n}}+O\left( \alpha ^{2}\right) .  \label{krr}
\end{eqnarray}

From Eq. (\ref{krr}) it is obvious that Kretschmann scalar diverges at $r=0$
and, like the asymptotically adS solutions, it reduces to $8\Lambda
^{2}/n(n+1)$\ for $r\longrightarrow \infty $.

Figs. \ref{metric1} - \ref{metric4} show that the singularity may be covered
with horizon and, therefore, we can interpret the singularity as a black
hole. In addition, these figures confirm that the nonlinearity parameter not
only modify the electromagnetic part of solutions, but also the kind of
horizons. For vanishing $\alpha$, with suitable choice of parameters, metric
function could acquire at most two horizons whereas, for this nonlinear
theory (nonzero $\alpha$), it is possible to find three horizons. Fig. \ref%
{metric4} indicates that the nonlinearity parameter may considerably affect
the existence, location and type of horizons. From Figs. \ref{metric1} - \ref%
{metric4} we find that, depending the values of metric parameters with
suitable $\alpha$, the horizons of the black hole solutions may be extreme
or not.

In order to study the conformal structure of the solutions, one may use the
conformal compactification method to plot the Carter-Penrose (conformal)
diagram (see Figs. \ref{Pen1}-\ref{Pen3}). The Carter-Penrose diagrams and
also the figures of the metric function (Figs. \ref{metric1} -\ref{metric3})
confirm that, the singularity is spacelike such as that of Schwarzschild
black holes. In other words, keeping the first order of nonlinearity
parameter and ignoring the higher order of $\alpha $, the timelike
singularity of the Reissner-Nordstr\"{o}m black holes (dotted line in Figs. %
\ref{metric1}-\ref{metric3}) change to a spacelike singularity. Drawing the
Carter-Penrose diagrams shows that the causal structure of the solutions are
asymptotically well behaved.

The temperature may be obtained through the use of regularity of the
solutions at $r=r_{+}$, yielding
\begin{equation}
T_{+}=\frac{f^{\prime }(r_{+})}{4\pi }=\frac{1}{2\pi \left( n-1\right)}
\left( \frac{\left( n-1\right) \left( n-2\right)k }{2r_{+}}-\Lambda r_{+}-%
\frac{q^{2}}{r_{+}^{2n-3}}+\frac{2q^{4}}{r_{+}^{4n-5}}\alpha \right)
+O\left( \alpha ^{2}\right) .  \label{T+}
\end{equation}

The electric potential $\Phi$, measured at infinity with respect to the
horizon, is defined by \cite{Cvetic,Caldarelli}
\begin{equation}
\Phi =A_{\mu }\chi ^{\mu }\mid _{r\rightarrow \infty }-A_{\mu }\chi ^{\mu
}\mid _{r=r_{+}},  \label{potential formula}
\end{equation}
with the following explicit form
\begin{equation}
\Phi =\frac{q}{\left( n-2\right) r_{+}^{n-2}}-\frac{4q^{3}}{\left(
3n-4\right) r_{+}^{3n-4}}\alpha +O\left( \alpha ^{2}\right) .
\label{potential}
\end{equation}

The last term in the right hand side of Eqs. (\ref{T+}) and (\ref{potential}%
) indicates the nonlinearity effect of the mentioned NLED.

%%%%%%%%%%%%%%%%%%%%%%%%%%%%%%%%%%%%%%%%%%%%%%%%%%%%%%%%%%%%%%%%%%%%%%%%%%%%%%%%%%%%%%%%%%%%%%%%%%%%%%%%%%%%%%%%%%%%%%%%%%%%%%%%%

\section{Thermodynamics of Asymptotically Flat Black Hole ($\Lambda =0$, $%
k=1 $) \label{K1}}

At first, we calculate the conserved and thermodynamic quantities of the
black hole for $\Lambda =0$ and $k=1$. Second, we obtain a Smarr-type
formula for the mass as a function of the entropy and the electric charge of
the solutions and finally check the first law of thermodynamics.

The first quantity which we are going to calculate is the entropy of the
black hole. More than thirty years ago, Bekenstein argued that the entropy
of a black hole in Einstein gravity is a linear function of the area of its
event horizon, which is the area law \cite{Bekenstein}. Therefore, the
entropy per unit volume $V_{n-1}$ of the presented black hole is equal to
one-quarter of the area of the horizon%
\begin{equation}
S=\frac{r_{+}^{n-1}}{4}.  \label{entropy}
\end{equation}

In order to obtain the electric charge per unit volume $V_{n-1}$ of the
black hole, we use the flux of the electric field at infinity, yielding%
\begin{equation}
Q=\frac{q}{4\pi },  \label{charge}
\end{equation}
which shows that, this kind of nonlinearity does not change the electric
charge. The ADM (Arnowitt-Deser-Misner) mass of black hole can be obtained
by using the behavior of the metric at large $r$. The mass per unit volume $%
V_{n-1}$ of the black hole is
\begin{equation}
M=\frac{\left( n-1\right) m}{16\pi },  \label{mass}
\end{equation}%
where we can obtain $m$\ from $f(r=r_{+})=0$.

After calculating all of the conserved and thermodynamic quantities of the
black hole solutions, we want to investigate the first law of
thermodynamics. To do this, we obtain the total mass $M$ as a function of
the extensive quantities $Q$\ and $S$. Using the expression for the entropy,
the electric charge and the mass given in Eqs. (\ref{entropy}), (\ref{charge}%
) and (\ref{mass}), and the fact that $f(r=r_{+})=0$, one can obtain a
Smarr-type formula as
\begin{equation}
M\left( S,Q\right) =\frac{\left( n-1\right) \Upsilon ^{n-2}}{16\pi }+\frac{%
2\pi Q^{2}}{\left( n-2\right) \Upsilon ^{n-2}}-\frac{64\pi ^{3}Q^{4}}{\left(
3n-4\right) \Upsilon ^{3n-4}}\alpha +O\left( \alpha ^{2}\right) ,
\end{equation}
where $\Upsilon =\left( 4S\right) ^{1/\left( n-1\right) }$.

Now, we regard the parameters $Q$ and $S$ as a complete set of extensive
parameters and define the intensive parameters conjugate to them. These
quantities are the temperature and the electric potential
\begin{equation}
T=\left( \frac{\partial M}{\partial S}\right) _{Q}=\frac{n\left( n-1\right)
\left( n-2\right) \Upsilon ^{n-2}}{16\pi n\left( n-1\right) S}-\frac{2\pi
Q^{2}}{S\left( n-1\right) \Upsilon ^{n-2}}+\frac{64\pi ^{3}Q^{4}}{S\left(
n-1\right) \Upsilon ^{3n-4}}\alpha +O\left( \alpha ^{2}\right),  \label{1}
\end{equation}
\begin{equation}
\Phi =\left( \frac{\partial M}{\partial Q}\right) _{S}=\frac{4\pi Q}{\left(
n-2\right) \Upsilon ^{n-2}}-\frac{256\pi ^{3}Q^{3}}{\left( 3n-4\right)
\Upsilon ^{3n-4}}\alpha +O\left( \alpha ^{2}\right) .  \label{2}
\end{equation}

Using Eqs. (\ref{entropy}) and (\ref{charge}), one can show that the Eqs. (%
\ref{1}) and (\ref{2}) are equal to Eqs. (\ref{T+}) and (\ref{potential}),
respectively. Thus, these quantities satisfy the first law of thermodynamics
\begin{equation}
dM=TdS+\Phi dQ.
\end{equation}

\subsection{Stability of the Solutions}

In what follows, we want to investigate the local stability of the charged
black hole solutions of Einstein gravity in the presence of nonlinear
electrodynamics in the canonical and the grand canonical ensembles. In
principle the local stability can be carried out by finding the determinant
of the Hessian matrix of $M(X_{i})$ with respect to its extensive variables $%
X_{i}$, $H_{X_{i}X_{j}}^{M}=\left[ \partial ^{2}M/\partial X_{i}\partial
X_{j}\right] $ \cite{Cvetic,Caldarelli}. In our case the mass $M$ is a
function of the entropy $S$\ and the charge $Q$. The number of thermodynamic
variables depends on the ensemble that is used. In the canonical ensemble,
the positivity of the heat capacity $C_{Q}=T_{+}/\left( \partial
^{2}M/\partial S^{2}\right) _{Q}$ is sufficient to ensure the local
stability. Since $T_{+}$ should be a positive definite quantity for physical
black holes, it is sufficient to check the sign of $\left( \partial
^{2}M/\partial S^{2}\right) _{Q}$
\begin{equation}
\left( \frac{\partial ^{2}M}{\partial S^{2}}\right) _{Q}=-\frac{(n-2)}{\pi
\left( n-1\right) r_{+}^{n}}+\frac{2\left( 2n-3\right) q^{2}}{\pi \left(
n-1\right) ^{2}r_{+}^{3n-4}}-\frac{4\left( 4n-5\right) q^{4}}{\pi \left(
n-1\right) ^{2}r_{+}^{5n-6}}\alpha +O\left( \alpha ^{2}\right) .  \label{CQ}
\end{equation}

%%%%%%%%%%%%%%%%%%%%%%%%%%%%%%%%%%%%%%%%%%%%%%%%%%%%%%%%%%
\begin{figure}[tbp]
\epsfxsize=10cm \epsffile{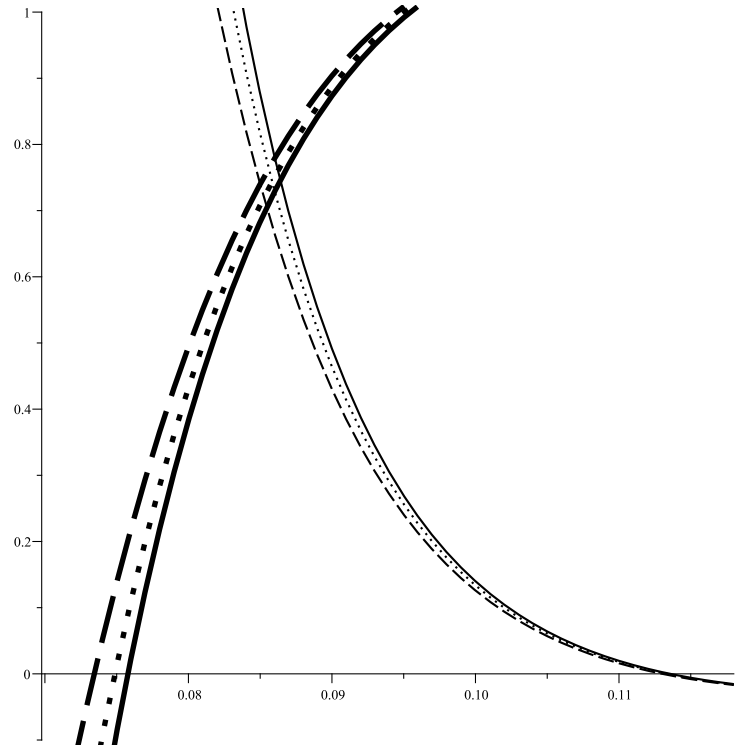}
\caption{\emph{Asymptotically flat solutions:} $10^{-4} \times\left( \frac{%
\partial ^{2}M}{\partial S^{2}}\right) _{Q}$ (thin lines) and $T_{+}$ (bold
lines) versus $r_{+}$ for $n=4$, $q=0.01$, $\protect\alpha=10^{-5}$ (solid
line), $\protect\alpha=5 \times 10^{-5}$ (dotted line) and $\protect\alpha%
=10 \times 10^{-5}$ (dashed line) }
\label{Stabilityalpha}
\end{figure}
%%%%%%%%%%%%%%%%%%%%%%%%%%%%%%%%%%%%%%%%%%%%%%%%%%%%%%%%%%

%%%%%%%%%%%%%%%%%%%%%%%%%%%%%%%%%%%%%%%%%%%%%%%%%%%%%%%%%%
\begin{figure}[tbp]
\epsfxsize=10cm \epsffile{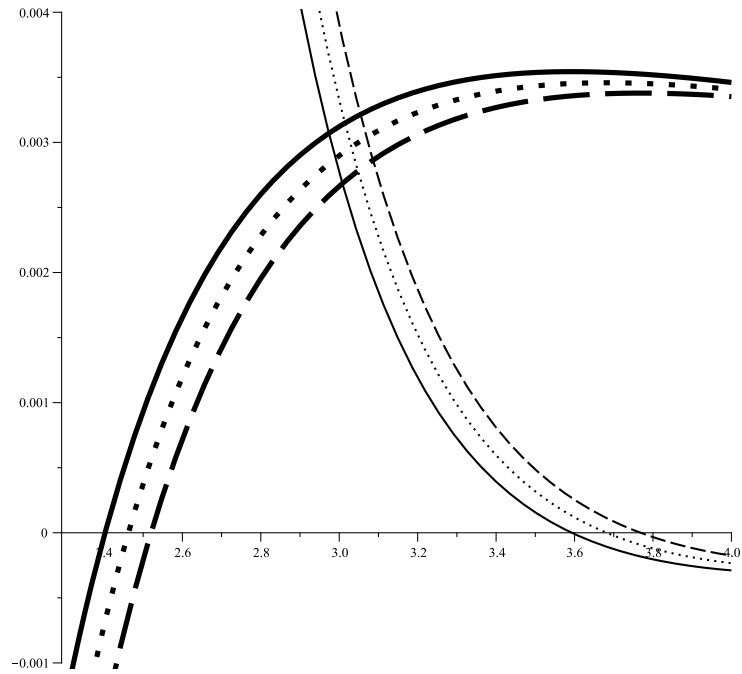}
\caption{\emph{Asymptotically flat solutions:} $\left( \frac{\partial ^{2}M}{%
\partial S^{2}}\right) _{Q}$ (thin lines) and $\frac{T_{+}}{10}$ (bold
lines) versus $r_{+}$ for $n=4$, $\protect\alpha=10^{-4}$, and $q=10$ (solid
line), $q=10.5$ (dotted line) and $q=11$ (dashed line) }
\label{Stabilityq}
\end{figure}
%%%%%%%%%%%%%%%%%%%%%%%%%%%%%%%%%%%%%%%%%%%%%%%%%%%%%%%%%%

Considering Eq. (\ref{CQ}), we find that the first and second terms are
related to the Einstein--Maxwell gravity and third one is related to the
effect of nonlinearity. In order to find the effects of nonlinearity on the
stability of the solutions, we plot Figs. \ref{Stabilityalpha} and \ref%
{Stabilityq}. These figures show that there is a lower limit, $r_{+min}$,
for the horizon radius of physical black holes (positive temperature). In
addition, considering Figs. \ref{Stabilityalpha} and \ref{Stabilityq}, one
finds large physical black holes are not stable. In other words, one can
obtain asymptotically flat stable black holes when the horizon radius
satisfies $r_{+min}<r_{+}<r_{+max}$, in which the values of $r_{+min}$ and $%
r_{+max}$ depend on $n$, $q$ and $\alpha $. Although Fig. \ref%
{Stabilityalpha} shows that decreasing $\alpha $ leads to increasing $%
r_{+min}$ (slightly increasing $r_{+max}$), Fig. \ref{Stabilityq} indicates
that decreasing $q$ leads to decreasing both $r_{+min}$ and $r_{+max}$.

In the grand canonical ensemble, after some algebraic manipulations, we
obtain
\begin{eqnarray}
H_{S,Q}^{M} &=&\frac{1}{\left( n-1\right) ^{2}}\left\{ -\frac{4\left[ \left(
n-1\right) \left( n-2\right) -2q^{2}r_{+}^{4-2n}\right] }{\left( n-2\right)
r_{+}^{2n-2}}+\right.   \nonumber \\
&&\left. \frac{16q^{2}\left[ 3\left( n-1\right) \left( n-2\right)
^{2}r_{+}^{4-4n}-\left( 7n-8\right) q^{2}r_{+}^{8-6n}\right] \alpha }{\left(
n-2\right) \left( 3n-4\right) }+O(\alpha ^{2})\right\} ,  \label{HSQ}
\end{eqnarray}%
where the last term is the nonlinearity effect of NLED. Regardless of the
values of $n$, $q$ and $\alpha $, we can write
\begin{eqnarray}
\left. H_{S,Q}^{M}\right\vert _{\text{Small }r_{+}} &=&-\frac{16\left(
7n-8\right) q^{4}\alpha }{\left( n-2\right) \left( 3n-4\right) r_{+}^{6n-8}}%
<0, \\
\left. H_{S,Q}^{M}\right\vert _{\text{Large }r_{+}} &=&-\frac{4}{\left(
n-1\right) r_{+}^{2n-2}}<0,
\end{eqnarray}%
which, in agreement with the canonical ensemble, confirm that the horizon
radius of asymptotically flat stable black holes should satisfy $%
r_{+min}<r_{+}<r_{+max}$.

%%%%%%%%%%%%%%%%%%%%%%%%%%%%%%%%%%%%%%%%%%%%%%%%%%%%%%%%%%%%%%%%%%%%%%%%%%%%%%%%%%%%%%%%%%%%%%%%%%%%%%%%%%%%%%%%%%%%%%%%%%%%%%%%%

\section{Thermodynamics of Asymptotically adS Rotating Black Branes with
Flat Horizon ($k=0$) \label{K0}}

Now, we want to endow our spacetime solutions (\ref{metric}) for $k=0$ with
global rotation parameters. In order to supplement angular momentum to the
spacetime, we perform the following rotation boost in the $t-\phi _{i}$
plane
\begin{equation}
\begin{array}{cc}
t\longmapsto \Xi t-a_{i}\phi _{i}, & \phi _{i}\longmapsto \Xi \phi _{i}-%
\frac{a_{i}}{l^{2}}t.%
\end{array}
\label{rotating}
\end{equation}

Thus the metric of $(n+1)$-dimensional asymptotically adS rotating spacetime
with $p$ rotation parameters can be written as
\begin{eqnarray}
ds^{2} &=&-f(r)\left( \Xi dt-\sum\limits_{i=1}^{p}a_{i}d\phi _{i}\right)
^{2}+\frac{r^{2}}{l^{4}}\sum\limits_{i=1}^{p}\left( a_{i}dt-l^{2}\Xi d\phi
_{i}\right) ^{2}+  \nonumber \\
&&\frac{dr^{2}}{f(r)}-\frac{r^{2}}{l^{2}}\sum\limits_{i=1}^{p}\left(
a_{i}d\phi _{j}-a_{j}d\phi _{i}\right)
^{2}+r^{2}\sum\limits_{i=p+1}^{n-1}dx_{i}^{2},  \label{rotating metric}
\end{eqnarray}%
where $\Xi =\sqrt{1+\sum_{i=1}^{p}a_{i}^{2}/l^{2}}$. Using Eq. (\ref{Maxwell
equation}), one can show that the suitable gauge potential can be written as%
\begin{equation}
A_{\mu }=h(r)\left( \Xi \delta _{\mu }^{0}-a_{i}\delta _{\mu }^{i}\right) \
\ (no\ sum\ on\ i),  \label{rotating vector potential}
\end{equation}%
where $h(r)$ is the same as that in Eq. (\ref{h(r)}). Now, we want to obtain
the metric function $f(r)$ for the spacetime (\ref{rotating metric}) by
inserting Eqs. (\ref{rotating metric}) and (\ref{rotating vector potential})
into Eq.\ (\ref{Field equation}). After some simplifications, we find that
the nonzero components of the gravitational field equations lead to four
different differential equations $e_{11}$, $e_{22}$, $e_{33}$ and $e_{44}$
for the unknown function $f(r)$, in which
\begin{eqnarray}
e_{11} &=&\left. e_{1}\left( static\ case\right) \right\vert _{k=0},
\label{e11} \\
e_{33} &=&\left. e_{2}\left(static\ case\right) \right\vert _{k=0},
\label{e22} \\
e_{33} &=&\left[ \left. e_{2}\left( static\ case\right) \right\vert _{k=0}%
\right] (\Xi ^{2}-1)-l^{2} f(r) \Xi ^{2}\left[ \left. e_{1}\left( static\
case\right) \right\vert _{k=0}\right] ,  \label{e33} \\
e_{44}&=&\sqrt{\Xi ^{2}-1} \{ \left[ \left. e_{1} \left(static\
case\right)\right\vert _{k=0}\right] \frac{f(r)}{2r^{2}}-\frac{1}{2l^{2}}%
\left[ \left. e_{2} \left(static\ case\right)\right\vert _{k=0}\right] \}.
\label{e44}
\end{eqnarray}
Considering Eqs. (\ref{e11}) and (\ref{e44}), we find that the metric
function (\ref{metric function}) with $k=0$ satisfies all field equations.
Straightforward calculations confirm that the mentioned rotating spacetime
has a curvature singularity at $r=0$, which may be covered with an event
horizon. We can obtain the temperature and the angular velocity of the event
horizon by analytic continuation of the metric function (\ref{rotating
metric}) and its regularity at the horizon $r_{+}$. One obtains
\begin{equation}
T_{+}=\left. \frac{T_{+}\left( static\ case\right) }{\Xi }\right\vert _{k=0},
\label{rotatingT+}
\end{equation}%
and
\begin{equation}
\Omega _{i}=\frac{a_{i}}{l^{2}\Xi }.  \label{angular velocity}
\end{equation}

Considering the fact that $\chi =\partial _{t}+\sum_{i=1}^{p}\Omega _{i}$ $%
\partial _{\phi _{i}}$ is the null Killing generator of the horizon and
using Eq. (\ref{potential formula}), one can find the electric potential as
\begin{equation}
\Phi = \frac{\Phi \left( static\ case\right)}{\Xi}.
\label{rotating potential}
\end{equation}

Here, we calculate other conserved and thermodynamic quantities of the black
brane solutions. Like previous section and with the same approaches, one can
show that the entropy and the electric charge per unit volume $V_{n-1}$ of
the presented black branes are, respectively, given by
\begin{equation}
S=\frac{\Xi r_{+}^{n-1}}{4},  \label{rotating entropy}
\end{equation}
and
\begin{equation}
Q=\frac{q\Xi }{4\pi }.  \label{rotating charge}
\end{equation}

Now, we should calculate the finite mass. In general, the action $I_{G}$,
diverges when evaluated on the solutions, as the Hamiltonian and other
associated conserved quantities. To compute the conserved charges of the
asymptotically adS solutions of Einstein gravity, we use the counterterm
method \cite{Kraus}. This method was inspired by the anti-de
Sitter/conformal field theory (AdS/CFT) correspondence and consists in
adding suitable counterterm $I_{ct}$ to the action $I_{G}$ in order to
ensure the finiteness of the boundary stress tensor derived by the
quasilocal energy definition \cite{Brown}. For asymptotically adS solutions
of Einstein gravity with flat boundary, the suitable counterterm $I_{ct}$ is
given by%
\begin{equation}
I_{ct}=-\frac{1}{8\pi }\int_{\partial \mathcal{M}}d^{n}x\sqrt{-\gamma }%
\left( \frac{n-1}{l}\right) .  \label{ct}
\end{equation}

Varying the total action ($I_{tot}=I_{G}+I_{ct}$) with respect to the
induced metric\ $\gamma _{ab}$, we find the boundary stress-tensor as%
\begin{equation}
T^{ab}=\frac{1}{8\pi }\left[ \Theta ^{ab}-\left( \Theta +\frac{n-1}{l}%
\right) \gamma ^{ab}\right] .  \label{stress-energy tensor}
\end{equation}

Now, we choose a spacelike surface $\mathcal{B}$ in $\partial \mathcal{M}$
with metric $\sigma _{ij}$, and write the boundary metric in
Arnowitt-Deser-Misner form%
\begin{equation}
\gamma _{\mu \nu }dx^{\mu }dx^{\nu }=-N^{2}dt^{2}+\sigma _{ij}\left(
d\varphi ^{i}+V^{i}dt\right) \left( d\varphi ^{j}+V^{j}dt\right) ,
\label{ADM form}
\end{equation}%
where the coordinates $\varphi ^{i}$ are the angular variables
parameterizing the hypersurface of constant $r$ around the origin, and $N$
and $V^{i}$ are the lapse and shift functions, respectively. When there is a
Killing vector field $\xi $ on the boundary, then the quasilocal conserved
quantities associated with the stress energy momentum tensor of Eq. (\ref%
{stress-energy tensor}) can be calculated as
\begin{equation}
Q\left( \xi \right) =\int_{\mathcal{B}}d^{n-2}\varphi \sqrt{\sigma }%
T_{ab}n^{a}\xi ^{b},  \label{conserved quantities}
\end{equation}%
where $\sigma $ is the determinant of the metric $\sigma _{ij}$, and $n^{a}$
is the timelike unit normal vector to the boundary $\mathcal{B}$. The
conserved quantities associated to the timelike $\xi =\partial _{t}$ and
rotational $\zeta =\partial _{\phi _{i}}$ Killing vector fields are
\begin{equation}
M=\int_{\mathcal{B}}d^{n-2}\varphi \sqrt{\sigma }T_{ab}n^{a}\xi ^{b}=\frac{%
V_{n-1}}{16\pi }m\left( n\Xi ^{2}-1\right) ,  \label{rotating mass}
\end{equation}

\begin{equation}
J_{i}=\int_{\mathcal{B}}d^{n-2}\varphi \sqrt{\sigma }T_{ab}n^{a}\zeta ^{b}=%
\frac{V_{n-1}}{16\pi }n\Xi ma_{i},
\end{equation}%
which are the mass and the angular momentum of the system enclosed by the
boundary $\mathcal{B}$. To check the first law of thermodynamics, we obtain
the total mass $M$ as a Smarr-type formula
\begin{equation}
M\left( S,J,Q\right) =\frac{\left( nZ-1\right) J}{nl\sqrt{Z\left( Z-1\right)
}},
\end{equation}%
where $J_{i}=\sum_{i}J_{i}^{2}$\ and $Z=\Xi ^{2}$ is the positive real root
of the following equation%
\begin{equation}
\frac{2\Lambda \Psi ^{-2}}{n\left( n-1\right) }+\frac{16\pi J\Psi ^{n-2}}{nl%
\sqrt{Z\left( Z-1\right) }}-\frac{32\pi ^{2}Q^{2}\Psi ^{2n-4}}{\left(
n-1\right) \left( n-2\right) Z}+\frac{1024\pi ^{4}Q^{4}\Psi ^{4n-6}}{\left(
n-1\right) \left( 3n-4\right) Z^{2}}\alpha +O\left( \alpha ^{2}\right) =0,
\end{equation}%
with $\Psi =\left[ \sqrt{Z}\left( 4S\right) ^{-1}\right] ^{1/(n-1)}$. It is
a matter of straightforward calculation to show that the conserved and
thermodynamic quantities satisfy the first law of thermodynamics
\begin{equation}
dM=TdS+\sum_{i}\Omega_{i} dJ_{i}+\Phi dQ.
\end{equation}

In other words, the quantities $T=\left( \frac{\partial M}{\partial S}%
\right) _{J,Q}$, $\Omega _{i}=\left( \frac{\partial M}{\partial J_{i}}%
\right) _{S,Q}$\ and $\Phi =\left( \frac{\partial M}{\partial Q}\right)
_{J,S}$\ are the same as those calculated in Eqs. (\ref{rotatingT+}), (\ref%
{angular velocity})\ and (\ref{rotating potential}), respectively.

%%%%%%%%%%%%%%%%%%%%%%%%%%%%%%%%%%%%%%%%%%%%%%%%%%%%%%%%%%
\begin{figure}[tbp]
\epsfxsize=10cm \epsffile{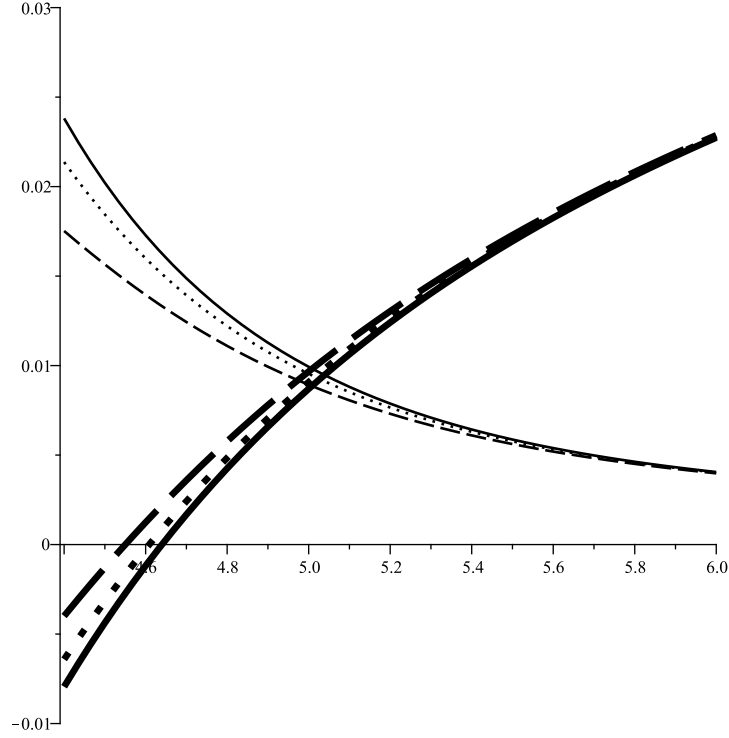}
\caption{\emph{Asymptotically adS solutions:} $\left( \frac{\partial ^{2}M}{%
\partial S^{2}}\right) _{Q,J}$ (thin lines) and $\frac{T_{+}}{10}$ (bold
lines) versus $r_{+}$ for $n=4$, $q=100$, $\Xi=1.1$ and $\Lambda=-1$, and $%
\protect\alpha=0.001$ (solid line), $\protect\alpha=0.02$ (dotted line) and $%
\protect\alpha=0.05$ (dashed line) }
\label{Stabilityalphak0}
\end{figure}
%%%%%%%%%%%%%%%%%%%%%%%%%%%%%%%%%%%%%%%%%%%%%%%%%%%%%%%%%%

%%%%%%%%%%%%%%%%%%%%%%%%%%%%%%%%%%%%%%%%%%%%%%%%%%%%%%%%%%
\begin{figure}[tbp]
\epsfxsize=10cm \epsffile{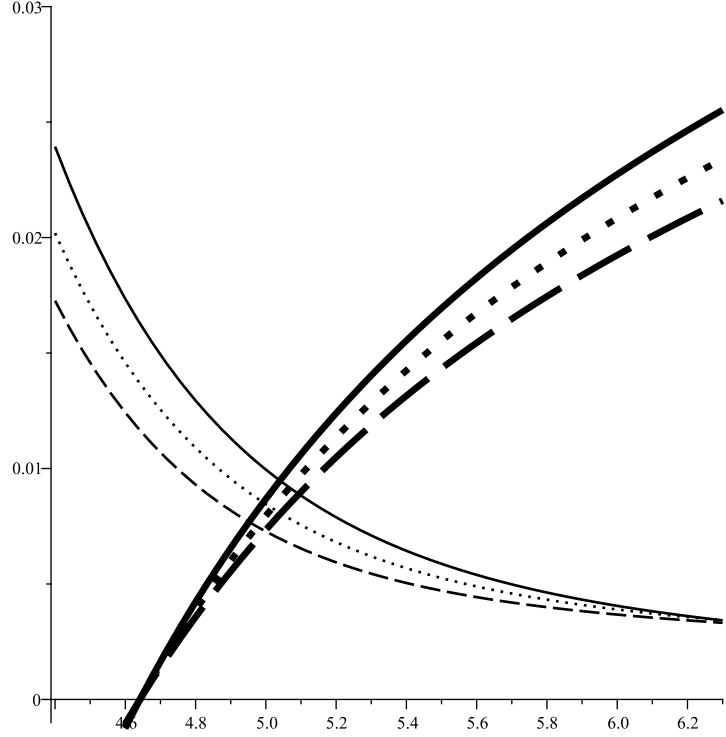}
\caption{\emph{Asymptotically adS solutions:} $\left( \frac{\partial ^{2}M}{%
\partial S^{2}}\right) _{Q,J}$ (thin lines) and $\frac{T_{+}}{10}$ (bold
lines) versus $r_{+}$ for $n=4$, $q=100$ and $\Lambda=-1$, $\protect\alpha%
=10^{-5}$ with $\Xi=1.1$ (solid line), $\Xi=1.2$ (dotted line) and $\Xi=1.3$
(dashed line) }
\label{StabilityXik0}
\end{figure}
%%%%%%%%%%%%%%%%%%%%%%%%%%%%%%%%%%%%%%%%%%%%%%%%%%%%%%%%%%

\subsection{Stability of the Solutions}

The final step is devoted to analyzing the local stability of charged
rotating black brane solutions of Einstein gravity in the presence of
nonlinear electrodynamics. We use the similar theoretical manner as was
discussed in the Sec. (\ref{K1}) and investigate thermal stability in both
the canonical and the grand canonical ensembles. In the canonical ensemble,
the electric charge and the angular momenta are fixed parameters, and $%
\left( \partial ^{2}M/\partial S^{2}\right) _{J,Q}$ at constant charge and
angular momenta is
\begin{equation}
\left( \frac{\partial ^{2}M}{\partial S^{2}}\right) _{J,Q}=\frac{C_{1}}{C_{2}%
}+\frac{C_{3}}{C_{4}}\alpha +O\left( \alpha ^{2}\right) ,  \label{CJQ}
\end{equation}
\begin{eqnarray}
C_{1} &=&-2\left\{ n\left[ 3(n-2)\Xi ^{2}-n+3\right] q^{4}r_{+}^{-3n+6}+%
\right.  \nonumber \\
&&2\left[ 3(n-2)\Xi ^{2}-n^{2}+3\right] \Lambda q^{2}r_{+}^{-n+4}+  \nonumber
\\
&&\left. (n-2)\left[ (n+2)\Xi ^{2}-n-1\right] \Lambda
^{2}r_{+}^{n+2}\right\} ,
\end{eqnarray}%
\begin{equation}
C_{2}=(n-1)\pi \Xi ^{2}\left[ (n-2)\Xi ^{2}+1\right] \left[ (n-2)\Lambda
r_{+}^{2n}-nq^{2}r_{+}^{2}\right] ,
\end{equation}%
\begin{eqnarray}
C_{3} &=&-4q^{4}\left\{ n^{2}\left[ (17n^{2}-54n+40)\Xi ^{2}-n(5n-23)-20%
\right] q^{4}r_{+}^{-5n+10}-\right.  \nonumber \\
&&2n(n-2)\left[ (13n^{2}-50n+40)\Xi ^{2}-n(n-19)-20\right] \Lambda
q^{2}r_{+}^{-3n+8}+  \nonumber \\
&&\left. (n-2)^{2}\left[ (5n^{2}-42n+40)\Xi ^{2}+n(7n+11)-20\right] \Lambda
^{2}r_{+}^{-n+6}\right\} ,
\end{eqnarray}%
\begin{equation}
C_{4}=(n-1)(3n-4)\pi \Xi ^{2}\left[ (n-2)\Xi ^{2}+1\right] \left[
(n-2)\Lambda r_{+}^{2n}-nq^{2}r_{+}^{2}\right] ^{2},
\end{equation}%
where in the Eq. (\ref{CJQ}) the first term is related to the
Einstein--Maxwell gravity.

Here, we plot Figs. \ref{Stabilityalphak0} and \ref{StabilityXik0} to
investigate the nonlinearity as well as rotation effects. These figures show
that for suitable fixed values of the metric parameters, there is an $%
r_{+min}$, in which for $r_{+}>r_{+min}$ we can obtain physical
asymptotically adS rotating stable black brane solutions. In other words, we
find that, unlike the asymptotically flat solutions, the asymptotically adS
rotating large black brane solutions are stable. Considering Fig. \ref%
{StabilityXik0}, one may find the rotation parameter affects on the values
of both $T_{+}$ and $\left( \frac{\partial ^{2}M}{\partial S^{2}}\right)
_{J,Q}$.

In the grand canonical ensemble, one can show that the determinant of the
Hessian matrix has the following form
\begin{equation}
H_{S,Q,J}^{M}=\frac{64\pi \left( q^{2}-\Lambda r_{+}^{2n-2}\right) }{B_{1}}-%
\frac{B_{2}}{B_{1}(3n-4)\left[ nq^{2}-(n-2)\Lambda r_{+}^{2n-2}\right]
r_{+}^{2n-2}}\alpha +O\left( \alpha ^{2}\right) ,  \label{HSQJ}
\end{equation}%
\begin{equation}
B_{1}=l^{2}\Xi ^{6}\left[ \left( n-2\right) \Xi ^{2}+1\right] \left[
nq^{2}-(n-2)\Lambda r_{+}^{2n-2}\right] r_{+}^{3n-4},
\end{equation}%
\begin{equation}
B_{2}=256\pi q^{2}\left[ 3(n-2)^{2}\Lambda
^{2}r_{+}^{4n-4}+3n(n-1)q^{4}-2(n-2)(3n-2)\Lambda q^{2}r_{+}^{2n-2}\right] ,
\end{equation}%
where in the Eq. (\ref{HSQJ}) the first term is the determinant of Hessian
matrix of Einstein--Maxwell gravity and the last term indicates the
nonlinearity effect. Following the method of previous section and regardless
of the values of $n$, $q$, $\Lambda $, $\Xi $ and $\alpha $, one finds
\begin{eqnarray}
\left. H_{S,Q,J}^{M}\right\vert _{\text{Small }r_{+}} &=&-\frac{768\pi
(n-1)q^{2}\alpha }{n(3n-4)l^{2}\Xi ^{6}\left[ \left( n-2\right) \Xi ^{2}+1%
\right] r_{+}^{5n-6}}<0, \\
\left. H_{S,Q,J}^{M}\right\vert _{\text{Large }r_{+}} &=&\frac{64\pi }{%
(n-2)l^{2}\Xi ^{6}\left[ \left( n-2\right) \Xi ^{2}+1\right] r_{+}^{3n-4}}>0,
\end{eqnarray}%
which are in agreement with results of the canonical ensemble and confirm
that large black branes are stable.

In order to analyze the correctness of our discussions for stability
criterion of asymptotically flat and adS black objects, we should argue for
the validity of numerical calculations. Regarding Eq. (\ref{E2(r)}), we find
that higher order terms of electric field can be formed by increasing $j$ in
$\left( \frac{q}{r_{+}^{n-1}}\right) ^{(2j+1)}\alpha ^{j}$ ($j=0,1,2,...$).
Therefore, ignoring the higher order terms of $\alpha $ makes sense, if
increasing $j$ leads to reasonable decreasing of $\left( \frac{q}{r_{+}^{n-1}%
}\right) ^{(2j+1)}\alpha ^{j}$. In the following tables, we consider the
numerical calculations of Figs. \ref{Stabilityalpha}--\ref{StabilityXik0}.

\begin{center}
\begin{tabular}{crc}
\hline\hline
Fig. \ref{Stabilityalpha} & \vline \vline \hspace{0.0001cm} \vline \vline &
Fig. \ref{Stabilityq} \\ \hline
$q=0.010$ & \vline \vline \hspace{0.0001cm} \vline \vline & $q=10.00$ \\
\hline
$\alpha=10^{-5}$ & \vline \vline \hspace{0.0001cm} \vline \vline & $%
\alpha=10^{-5}$ \\ \hline
$r_{+}\approx 0.075$ & \vline \vline \hspace{0.0001cm} \vline \vline & $%
r_{+}\approx2.40$ \\ \hline
$\frac{q}{r_{+}^{3}}\approx 23.704$ & \vline \vline \hspace{0.0001cm} \vline %
\vline & $\frac{q}{r_{+}^{3}}\approx 0.72$ \\ \hline
$\left( \frac{q}{r_{+}^{3}}\right) ^{3} \alpha \approx 0.133$ & \vline %
\vline \hspace{0.0001cm} \vline \vline & $\left( \frac{q}{r_{+}^{3}}%
\right)^{3} \alpha \approx 0.38 \times 10^{-5}$ \\ \hline
$\left( \frac{q}{r_{+}^{3}}\right) ^{5} \alpha^2 \approx 0.75 \times 10^{-3}$
& \vline \vline \hspace{0.0001cm} \vline \vline & $\left( \frac{q}{r_{+}^{3}}
\right) ^{5} \alpha^2 \approx 0.20 \times 10^{-10}$ \\ \hline
\end{tabular}
\hspace{2cm}
\begin{tabular}{crc}
\hline\hline
Fig. \ref{Stabilityalphak0} & \vline \vline \hspace{0.0001cm} \vline \vline
& Fig. \ref{StabilityXik0} \\ \hline
$q=100$ & \vline \vline \hspace{0.0001cm} \vline \vline & $q=100$ \\ \hline
$\alpha=10^{-3}$ & \vline \vline \hspace{0.0001cm} \vline \vline & $%
\alpha=10^{-5}$ \\ \hline
$r_{+}\approx 4.6$ & \vline \vline \hspace{0.0001cm} \vline \vline & $%
r_{+}\approx 4.64$ \\ \hline
$\frac{q}{r_{+}^{3}}\approx 1.03$ & \vline \vline \hspace{0.0001cm} \vline %
\vline & $\frac{q}{r_{+}^{3}}\approx 1.00$ \\ \hline
$\left( \frac{q}{r_{+}^{3}}\right) ^{3} \alpha\approx 10^{-3}$ & \vline %
\vline \hspace{0.0001cm} \vline \vline & $\left( \frac{q}{r_{+}^{3}}%
\right)^{3} \alpha \approx 10^{-5}$ \\ \hline
$\left( \frac{q}{r_{+}^{3}}\right) ^{5} \alpha^2 \approx 10^{-6}$ & \vline %
\vline \hspace{0.0001cm} \vline \vline & $\left( \frac{q}{r_{+}^{3}} \right)
^{5} \alpha^2 \approx 10^{-10}$ \\ \hline
\end{tabular}
\\[0pt]
\vspace{0.1cm} Left table: asymptotically flat solutions \& Right
table: asymptotically adS solutions. \vspace{0.5cm}
\end{center}

Taking into account the numerical results of the tables, one can confirm
that numerical calculations of stability conditions for both asymptotically
flat and adS black objects are logical.

%%%%%%%%%%%%%%%%%%%%%%%%%%%%%%%%%%%%%%%%%%%%%%%%%%%%%%%%%%%%%%%%%%%%%%%%%%%%%%%%%%%%%%%%%%%%%%%%%%%%%%%%%%%%%%%%%%%%%%%%%%%%%%%%%

\section{Conclusions \label{Conclusions}}

Motivated by the (quartic) string corrections of Maxwell field
strength, at first, we obtained black hole solutions of
Einstein-NLED gravity with various horizon topology and
investigated their geometric properties. Then, we fixed $\Lambda
=0$ and $k=1$ to calculate the conserved quantities of the
asymptotically flat black holes. We obtained a Smarr-type formula
for the mass as a function of the entropy and the electric charge
of the solutions and checked the first law of thermodynamics. We
studied the stability analysis of the asymptotically flat black
holes both in the canonical and the grand canonical ensembles and
investigated the effects of NLED. We found that for the fixed
values of $n$, $q$ and $\alpha $, small and large physical black
holes are not stable. It means that obtained asymptotically
flat black holes can be stable when the horizon radius satisfies $%
r_{+min}<r_{+}<r_{+max}$, in which the values of $r_{+min}$ and $r_{+max}$
depend on the metric parameters. In addition, we found that although
decreasing the nonlinearity parameter leads to increasing $r_{+min}$
(slightly increasing $r_{+max}$), decreasing the charge parameter leads to
decreasing both $r_{+min}$ and $r_{+max}$.

After that, we considered the horizon-flat solutions and used a suitable
rotation boost to endow angular momentum to the asymptotically adS
spacetime. Using the counterterm method, we obtained the conserved
quantities of the asymptotically adS black branes. We also obtained a
Smarr-type formula for the finite mass as a function of the other quantities
and showed that they satisfy the first law of thermodynamics. Besides, we
performed a stability analysis of the rotating solutions both in the
canonical and the grand canonical ensembles. We showed that there is a lower
limit, $r_{+min}$, for the physical solutions (positive temperature).
Stability analysis of both ensembles confirmed that, unlike the
asymptotically flat solutions with spherical horizon, the horizon-flat
asymptotically adS rotating black brane solutions with large event horizon
are stable. In other words, we showed that there is an $r_{+min}$ for
suitable fixed values of the metric parameters, in which for $r_{+}>r_{+min}$%
, the asymptotically adS rotating black brane solutions are stable.
Moreover, we fixed the values of $n$, $\Lambda $, $\alpha $ and $q$ to
analyze the rotation's effect on the stability conditions. We showed that
although $\Xi $ does not change the location of $r_{+min}$, it can change
the values of the temperature, the heat capacity and the determinant of
Hessian matrix.

It is notable that due to the negative temperature, the small black
holes/branes are not physical and we should restrict the horizon radius to $%
r_{+}>r_{+min}$, while (in)stability of large ones is related to their
horizon geometries. In other words, large black holes (branes) with $k=1$ ($%
k=0$) are unstable (stable).

Finally, it was seen that the nonlinearity part not only modified
electromagnetic part of the solutions, but also the kind of
horizons and thermodynamics properties. In absence of of
correction part, with suitable choices of parameters, metric
function could acquire two horizons whereas, for this nonlinear
theory, it is possible to find three horizons. This fact has some
application regarding anti-evaporation of black holes/branes. The
structure of black hole in presence of nonlinear electrodynamics
is quite different comparing to the linear Maxwell theory and its
phenomenology is also describing a more general case. In addition,
it is worthwhile to mention that considering the first order
effects of nonlinear electrodynamics changed the properties of the
black objects at small distances. In other words, this
generalization changed timelike singularity of
Reissner-Nordstr\"{o}m black holes to spacelike singularity.
Hence, in order to recover the properties of the
Reissner-Nordstr\"{o}m black holes, it may be logical to keep
terms only upto quadratic order of $\alpha $ in the series
expansions. Besides, it is
notable that one can consider asymptotically adS black holes with spherical $%
(k=1)$ and hyperbolic topologies $(k=-1)$, to investigate $P-V$ criticality
in the extended phase space of the solutions by calculating the Gibbs free
energy for various $\Lambda $. These extensions are under examination.

\begin{acknowledgements}
We would like to thank the anonymous referee for useful
suggestions and enlightening comments. We also thank S. Panahiyan
and B. Eslam Panah for reading the manuscript. We wish to thank
Shiraz University Research Council. This work has been supported
financially by the Research Institute for Astronomy and
Astrophysics of Maragha, Iran.
\end{acknowledgements}

\end{document}